\documentclass[epjCONF]{svjour}
\usepackage{graphicx}
\usepackage[varg]{txfonts} 
\usepackage[latin1]{inputenc}
\usepackage{wrapft}
\session-title{MESON2012 - 12th International Workshop on Meson Production, Properties and Interaction}
\begin{document}
\title{$\eta^{\prime}$ meson under partial restoration of chiral symmetry in nuclear medium\thanks{This work was partially supported by the Grants-in-Aid for Scientific Research (No. 22740161). This work was done in part under the Yukawa International Program for Quark- hadron Sciences (YIPQS).}
}
\author{Daisuke Jido\inst{1}\fnmsep\thanks{\email{jido@yukawa.kyoto-u.ac.jp}} \and Hideko Nagahiro\inst{2} \and Satoru Hirenzaki\inst{2} \and Shuntaro Sakai\inst{3} }
\institute{Yukawa Institute for Theoretical Physics, Kyoto University, Kyoto 606-8502, Japan \and Department of Physics, Nara Women's University, 
Nara 630-8506, Japan \and Department of Physics, Graduate School of Science, Kyoto University, Kyoto 606-8502, Japan}
\abstract{
We shed light upon the $\eta^{\prime}$  mass in nuclear matter in the context of partial restoration of chiral symmetry, pointing out that the U$_{A}$(1) anomaly effects causes the $\eta^{\prime}$-$\eta$ mass difference necessarily through the chiral symmetry breaking. As a consequence, it is expected that the $\eta^{\prime}$ mass is reduced by order of 100 MeV in nuclear matter where partial restoration of chiral symmetry takes place. The  discussion given here is based on Ref.~\cite{Jido:2011pq}
} 
%


\maketitle
%

\section{Introduction}
\label{intro}
Substantial description of hadron properties in terms of quark-gluon language brings us not only deeper insight of hadron but also its systematic understanding. Partial restoration of chiral symmetry in nuclei has been phenomenologically confirmed in pionic atom observation and low-energy pion-nucleus scattering with help of theoretical examination~\cite{Jido:2008bk}. The current issues on partial restoration of chiral symmetry in nuclear matter is precise determination of the density dependence of the quark condensate and systematic confirmation of the partial restoration by observing it in other systems. 

The $\eta^{\prime}$ meson is such a mysterious to have a larger mass than other light pseudoscalar mesons.  The U$_{A}$(1) anomaly effect is responsible for the large mass. It is important noting that the U$_{A}$(1) anomaly can affect on the $\eta^{\prime}$ mass only through the dynamical and/or explicit breaking of the SU(3) chiral symmetry~\cite{Lee:1996zy,Jido:2011pq}. This implies that the chiral symmetry breaking induces the mass splitting of the $\eta$ and $\eta^{\prime}$ mesons through the presence of the U$_{A}$(1) anomaly. Therefore, since the partial restoration of chiral symmetry takes place in nuclear matter, one expects a strong mass reduction of the $\eta^{\prime}$ meson there~\cite{Jido:2011pq}. Such a strong mass reduction in nuclear matter corresponds to a strong attractive force between the $\eta^{\prime}$ meson and a finite nucleus. With strong attraction one expects to have $\eta^{\prime}$ bound states in nuclei, which can be studied by formation experiments of  $\eta^{\prime}$-nucleus bound systems~\cite{Nagahiro:2004qz,Nagahiro:2006dr,Nagahiro:2011fi,Itahashi:2012ut}.


\section{Chiral property of the $\eta^{\prime}$ meson}

The key of the present discussion is the chiral symmetry property of the $\eta^{\prime}$ meson. The QCD Lagrangian has a chiral U$_{L}(3)\times$U$_{R}(3)$ symmetry in the chiral limit as quark flavor rotations. But U$_{A}$(1) is broken by quantum effect known as the U$_{A}$(1) anomaly. Since the U$_{V}$(1) symmetry trivially gives baryon number conservation, we concentrate on the chiral SU$_{L}(3)\times$SU$_{R}(3)$ symmetry. For simplicity, we assume that the U$_{A}$(1) symmetry is always broken due to the anomaly.\footnote{The U$_{A}$(1) anomaly itself can change in finite temperature and density systems due to the reduction of the instanton density there~\cite{Pisarski:1983ms}. Such cases are also included in the present discussion. 
} 
We also assume the flavor SU(3) symmetry in which there is no single-octet mixing and $\eta^{\prime}$ is a purely flavor singlet particle $\eta_{0}$.

\begin{wrapfigure}{r}{0.45\columnwidth}
\begin{center}
\resizebox{0.45\columnwidth}{!}{%
  \includegraphics[bb=0 0 872 524]{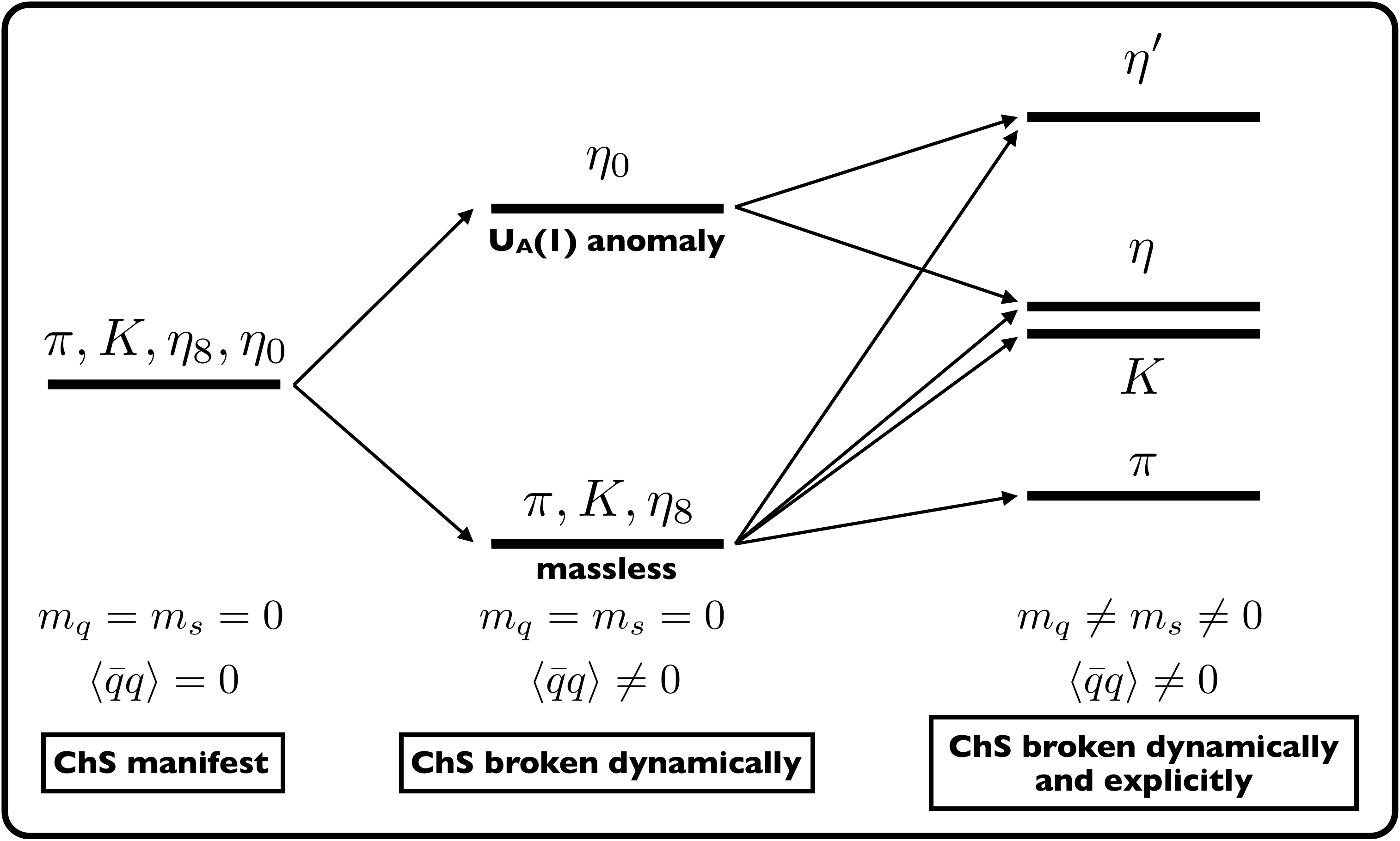}}
\end{center}
\caption{Light pseudoscalar meson spectrum in the various patterns of the SU(3) chiral symmetry breaking. In the left,  chiral symmetry is manifest without explicit nor dynamical breaking. All the pseudoscalar mesons have a same mass. In the middle, chiral symmetry is dynamically broken in the chiral limit.
The octet pseudoscalar mesons are identified as the Nambu-Goldstone bosons associated with the symmetry breaking.  In the right, chiral symmetry is broken dynamically by the quark condensate and explicitly by  finite quark masses. }
\label{fig:1}       
\end{wrapfigure}
%

Now let us consider the chiral multiplet of the pseudoscalar mesons. The simplest quark configuration of the pseudoscalar mesons is $\bar q^{i}_{L} q^{j}_{R} - \bar q^{j}_{R} q^{i}_{L}$, where $i$ and $j$ are flavor indices. Since quark and antiquark belong to the fundamental representations, $\bf 3$ and $\bf \bar 3$, respectively, the pseudoscalar mesons belong to the chiral representation $\bf (\bar 3,3) \oplus (3,\bar 3)$. In this chiral multiple we have also the scalar mesons which are the chiral partners of the pseudoscalar mesons. In this representation there are $3 \times 3 + 3\times 3 =18$ particles, 9 scalar  and 9 pseudoscalar mesons. When chiral symmetry is restored in the chiral limit, these 18 particles should degenerate with having a same mass. 
These 18 particles can be classified in terminology of the flavor SU$_{V}$(3). Since SU$_{V}$(3) does not distinguish the left and right chiralities, and if we recall $\bf \bar 3 \otimes 3 = 1 \oplus 8$, we find two sets of the flavor singlet and octet mesons,
that is, $\pi,K,\eta_{8}$ as octet pseudoscalar, $\eta_{0}$ as singlet pseudoscalar, $a_{0},\kappa, f_{0}$ as octet scalar and $\sigma$ as singlet scalar. Therefore, if chiral symmetry is manifest without explicit nor dynamical breaking, these 18 particles should have a same mass. This means especially no mass difference between $\eta_{8}$ and $\eta_{0}$ in a chiral restoration limit\footnote{In the flavor SU(2) case, even if chiral symmetry is manifest, $\pi$ and $\eta$ do not degenerate, since they belong to different representations in the SU(2) chiral
group. This fact can be understood as follows. The SU(2) is a special case of SU(3) with  symmetry breaking by the strange quark. Thus in the SU(2) case, $\eta_{0}$ and $\pi$ do not have to degenerate due to the SU(3) symmetry breaking.}, even though the U$_{A}$(1) symmetry is broken. 

When chiral symmetry is dynamically broken by the quark condensate, only the octet pseudoscalar mesons become massless since they are the Nambu-Goldstone bosons associated with the symmetry breaking. (Recall that the broken generators in SU$_{L}(3)\times$SU$_{R}(3)$ belong to the octet representation of the flavor SU(3).) Since $\eta_{0}$ is not a Nambu-Goldstone boson, it can be massive. In other words, the U$_{A}$(1) effect is turned on by chiral symmetry breaking and lifts the $\eta_{0}$ mass. In Fig.~\ref{fig:1} we show a schematic view of the pseudoscalar meson spectra in various chiral symmetry breaking patterns. 

\section{The $\eta^{\prime}$ mass in nuclear matter}

As discussed in the previous section, even though in the presence of the U$_{A}$(1) anomaly, when chiral symmetry is restored in the chiral limit, $\eta$ and $\eta^{\prime}$ should degenerate. This implies that the mass difference between $\eta$ and $\eta^{\prime}$ may be controlled by dynamical breaking of chiral symmetry. Now let us assume that the mass difference $\eta$ and $\eta^{\prime}$ comes from the anomaly effect and  the mass difference is proportional to the quark condensate. In this picture, since partial restoration of chiral symmetry takes place at the saturation density $\rho_{0}$ with about 35 \% reduction of the quark condensate in magnitude~\cite{Jido:2008bk}, the $\eta$-$\eta^{\prime}$ mass difference is suppressed at $\rho=\rho_{0}$ by 150 MeV. Recalling that the $\eta$ meson is one of the Nambu-Goldstone bosons, we expect that in-medium mass modification of the $\eta$ meson is small. Thus, the $\eta^{\prime}$ meson will get a strong attraction of order of 150 MeV at the saturation density. This scenario can be checked by chiral effective theories, such as NJL model~\cite{Nagahiro:2006dr} or linear sigma model~\cite{Sakai}.

Now, in order to discuss $\eta^{\prime}$-nucleus bound systems with this strong attraction, let us take a simple $\eta^{\prime}$ optical potential for finite nuclei
\begin{equation}
  V_{\eta^{\prime}}(r) = V_{0} \frac{\rho(r)}{\rho_{0}}
\end{equation}
where we assume the Woods-Saxon form of the nuclear density distribution $\rho(r)$. The strength of the potential are parameters. Here we take from 100 to 200 MeV for the real part of $V_{0}$, since the $\eta^{\prime}$ mass reduction at $\rho=\rho_{0}$ is expected to be 150 MeV. 
For the imaginary part, we take 20 MeV, which corresponds to 40 MeV absorption width in nuclear matter. This value is larger than the experimental value extracted from the transparency ratio in the $\gamma A \to \eta^{\prime} X$ reaction, 15-25 MeV~\cite{Nanova:2012vw}.
 


\begin{wrapfigure}{r}{0.38\columnwidth}
\begin{center}
\resizebox{0.38\columnwidth}{!}{%
  \includegraphics[bb=20 33 515 343]{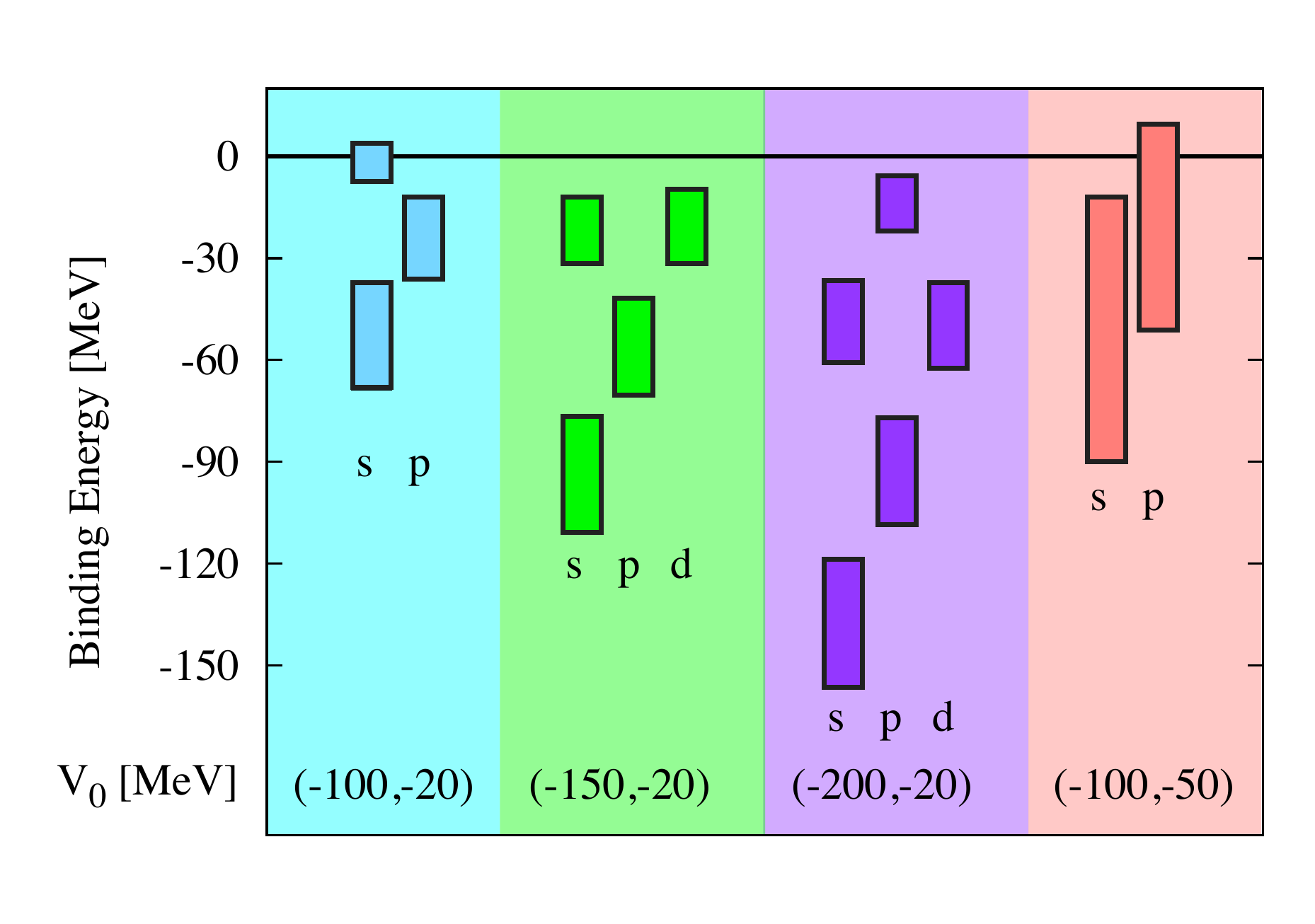}}
\end{center}
\caption{Bound state spectra of $\eta^{\prime}$-$^{11}$C.}
\label{fig:2}       
\end{wrapfigure}

In Fig.~\ref{fig:2} we show expected spectra of the $\eta^{\prime}$ bound states in $^{11}$C with several sets of the potential parameter $V_{0}$. Thanks to weak absorption  and strong attraction, the bound states are well separated except for the case of a large absorption width. This gives big advantage for formation experiments of $\eta^{\prime}$ nucleus bound systems to observe clearer peak structure. 

It is an interesting issue to see how strong the elementary $\eta^{\prime}N$ interaction is under having such strong attraction in nuclear matter. For this purpose, one needs chiral effective theories having $\eta^{\prime}$ and $N$ in the same footing and the mechanism of partial restoration of chiral symmetry at finite nuclear density. For example, two of the authors find using a linear sigma model with the flavor SU(3) breaking
that the $s$-wave $\eta^{\prime}N$ interaction can be similar in strength to the attraction in $\bar KN$ with $I=0$ obtained by the Tomozawa-Weinberg interaction at the $\bar KN$ threshold energy~\cite{Sakai}. 
Thus, we could have a bound state in a two-body $\eta^{\prime}N$. Nevertheless, phenomenological observation of the $\eta^{\prime}N$ scattering length which has been extracted from the three-body $pp\eta^{\prime}$ final state would be suggested to be as small as a few 0.1 fm with an unknown sign~\cite{Moskal:2000gj}. The small scattering length would indicate weak $\eta^{\prime}N$ interaction~\cite{Oset:2010ub}. If this would be the case, since the strong attraction discussed above comes from scalar exchange, one could expect some strong vector repulsion to have the $\eta^{\prime}N$ interaction weak. But there is no Tomozawa-Weinberg interaction, which is derived by vector meson exchange, between $\eta^{\prime}$ and $N$. Thus, any further experimental data are important for the understanding of the $\eta^{\prime}N$ interaction and in-medium $\eta^{\prime}$ properties.

\section{Conclusion}
We point out that partial restoration of chiral symmetry in a nuclear medium induces suppression of the U$_{A}$(1) anomaly effect to the $\eta^{\prime}$ mass. Consequently, we expect a large mass reduction of the $\eta^{\prime}$ meson in nuclear matter with relatively smaller absorption. The mass reduction can be observed as $\eta^{\prime}$-nucleus bound states in the formation reactions. The interplay between chiral symmetry restoration and the U$_{A}$(1) anomaly effect can be a clue to understand the $\eta^{\prime}$ mass generation mechanism. Therefore, experimental observations of deeply $\eta^{\prime}$-nucleus bound states, or even confirmation of nonexistence of such deeply bound states, is important to understand the U$_{A}$(1) anomaly effects on hadrons.


\end{document}